\documentclass[twocolumn]{aastex631}

\newcommand{\Msun}{$M_\odot$} 
\newcommand{\fsecs}{\mbox{$.\!\!^{s}$}}
\newcommand{\degs}{\mbox{$^{\circ}$}}

\newcommand{\axaf}{\mbox{\em Chandra\/}}

\usepackage{graphicx}

\usepackage{color}
\usepackage{datetime}

\shorttitle{Milli-arcsec X-ray astrometry of HE 0435$-$1223}
\shortauthors{Rogers, Schwartz, Spingola, \& Barnacka}

\begin{document}
\title{Milli-arcsec X-ray positions and X-ray varstrometry for the strongly  lensed AGN HE 0435-1223}

\correspondingauthor{Daniel Schwartz}
\email{das@cfa.harvard.edu}

\author[0009-0009-2028-6521]{Alysa Rogers}
\affiliation{Physics Department, Brandeis University, Waltham, MA 02451, USA}
\affiliation{Physics Department\footnote{Present Address}, University of Wisconsin–Madison, Madison, WI 53706, USA}

\author[0000-0001-8252-4753]{Daniel Schwartz}
\affiliation{Smithsonian Astrophysical Observatory, Cambridge, MA 02138, USA}

\author[0000-0002-2231-6861]{Cristiana Spingola}
\affiliation{INAF $-$ Istituto di Radioastronomia, Via Gobetti 101, I$-$40129, Bologna, Italy}

\author[0000-0001-5655-4158]{Anna Barnacka}
\affiliation{Smithsonian Astrophysical Observatory, Cambridge, MA 02138, USA}
\affiliation{Astronomical Observatory, Jagiellonian University, ul. Orla 171, 30-244 Cracow, Poland}

\begin{abstract}

Active galactic nuclei (AGN) are some of the most powerful objects in the Universe. For this reason, they can be observed up to high redshifts ($z$), giving valuable  insights into the evolution of our Universe.
However, high-$z$ AGN  are too distant to be spatially resolved with current or upcoming X-ray facilities.  In this paper we show how we can exploit gravitationally lensed AGN to significantly increase spatial resolution even at high-$z$. We combine astrometric data from \textit{Gaia} DR3 with imaging from the \textit{Chandra} X-ray Observatory of the quadruply-lensed quasar HE 0435--1223 to measure for the first time possible offsets between the optical and the X-ray emissions. We measure the X-ray source position for HE 0435-1223 within a 1$\sigma$ quasi-elliptical region of 0.5$\times$1.3 milli-arcsecond (mas), about 150 pc$^2$ at the redshift of the source ($z=1.689$). We find evidence for the X-ray emission being offset by a projected 3 mas from the  \textit{Gaia} (optical) emission. The positional offset is most likely associated to a portion of the X-ray emission arising from an X-ray jet or outflow. We also discuss how this method can be used to indicate the presence of a binary/offset AGN system. 

\end{abstract}


\section{Introduction} \label{sec:intro}

Active galactic nuclei (AGN) produce a significant portion of the total electromagnetic energy in the Universe. This emission comes from supermassive black holes (SMBHs) at the center of galaxies, produced by the conversion of gravitational potential energy to kinetic energy and electromagnetic radiation as matter falls toward the SMBH. In addition to the radiation, some of the matter in the accretion disk surrounding the black hole can escape as a collimated jet or uncollimated wind outflow \citep{Hardcastle2020}. 

The AGN emission covers the broad electromagnetic spectrum. The standard paradigm \citep{Urry1995} models the accretion disk emitting in optical and UV 
mostly at $10^{14-15.5}$ cm \citep[e.g.,][]{Morgan2010}  for a 10$^8$\Msun\,  SMBH, an obscuring dust torus re-emitting absorbed radiation in the IR band at a distance of $\sim 10^{17}$ cm \citep[e.g.,][]{Netzer2015}, and a hot corona  emitting in X-rays around the $\sim 10^{13.5}$ cm  \citep[e.g.,][]{Chartas2016} black hole and the disk. In at least $10$\% of the entire AGN population, there are also collimated jets that emit from radio to gamma wavelengths.  The observed properties  depend on accretion rate into the black hole, orientation relative to the observer, the presence (or not) of a relativistic jet, and potentially the host galaxy and environment \citep{Padovani2017}. For distant AGN, most physical regions cannot be directly imaged at wavelengths shortward of the radio and sub-mm emission. Additionally, the bolometric radiative luminosity, the energy outflow in radio jets, and the black hole mass must often be deduced via proxy measurements and models.

Although we have detailed empirical understanding of the multi-wavelength properties of AGN, their formation and evolution process is still unclear and debated. 
The categories of theories for initial formation of  black holes include the collapse of a massive star, dynamical evolution of dense nuclear star clusters, and collapse of a metal-free gas cloud (see \citealt{Volonteri2021Nature} for a comprehensive recent review). Whatever their formation process, as galaxies merge, 
dual (separated by $<10$ kpc), binary (separated by $<100$ pc) or offset AGN systems should be formed. However, they are observationally challenging to identify and characterize, hence, currently, their occurrence is under constrained especially at high-z \citep{Spingola2022}.

X-ray emission is a ubiquitous indicator of SMBH activity, and astrometric measurement of the X-ray structure potentially can reveal SMBH pairs, offset SMBH, X-ray location in comparison to the optical/radio emission, or X-rays from jets or mass outflows. Therefore, X-rays may be the most efficient band for finding such sources (see also \citealt{DeRosa2019, DeRosa2023}).
However, at X-ray wavelengths where some of the most energetic processes take place, angular resolution is limited to $\approx$0\farcs5 as achieved by the \mbox{\em Chandra X-ray Observatory\/}. Planned X-ray observatories for even the next several decades will not have substantially better angular resolution \citep{Gaskin2018,Mushotzky2018}, so a 10 pc separation pair can be resolved only nearer than about 10 Mpc. 

Gravitational lensing can greatly extend the spatial resolution of X-ray imaging \citep{Barnacka2017,Barnacka2018}. \citet{Spingola2020} verified this capability by comparing the radio and optical location of two quasars at redshift $\sim$1.4. We have been exploring the application of gravitational lensing for locating the of X-ray emission relatively to optical and radio emission \citep{Schwartz2021,Spingola2022}.  \citet{Barnacka2017,Barnacka2018} has 
demonstrated the use of the amplification by lensing to probe multi-wavelength relationships on angular scales not otherwise accessible \citep{Barnacka2015b, Barnacka2016, Spingola2020, Schwartz2021, Spingola2022}. As distinct from a refracting lens in visible light, the physics of a gravitational lens is strictly achromatic\footnote{Microlensing, absorption in the lensing galaxy, and source structure can all change the relative image fluxes, but do not perturb the position of a given emission region by more than a microarcsecond.} but is also highly astigmatic. We can take advantage of the first property to reference the X-ray images to the optical images. As long as we assume the same lens model for the optical and X-ray bands, then the predicted positions when mapping from the source plane to the image plane must be the same if and only if the emission is from the same source position. We will be modeling the emission as a point source in both bands, as that is all the observations reveal. Thus we do not concentrate on obtaining the "best" possible lens mass model, but only in predicting the optical positions to much better than the raw X-ray position locations. 

Barnacka (2017, see Fig. 3) shows how the spatial amplification ratio can approach the flux magnification ratio for sources separated in a direction perpendicular to the caustic, while very much less amplified at small angles relative to the caustic tangent. This distortion is well known from observations of the large arcs observed from lensed galaxies. 
Because image position changes so drastically near the caustics, spatial structure of a source is most easily measured for a system located close to and inside the caustic. We furthermore need sources inside the caustic because we need four images to be produced to over-determine a relative reference frame. The method used in this project is powerful because we use strong lensing and our analysis is based solely on image position. The position of images is least affected by external factors, whereas magnification and time delay can be impacted by microlensing and by absorption in the intervening galaxy.

\subsection{The gravitationally lensed quasar HE 0435-1223} \label{sec:style}

This work focuses on the optical and X-ray source emission in the lensed AGN  HE 0435--1223 (hereafter abbreviated HE0435, \citealt{Wisotzki2000, Wisotzki2002}). 

The background source is an AGN at a redshift $z_s = 1.693$ \citep{Sluse2012} strongly lensed into four images by a galaxy at $z_l = 0.4546 \pm 0.0002$, which belongs to a group of 12 galaxies \citep{Morgan2005, Eigenbrod2006b}. The SMBH mass of the background source has been estimated to be 2.3$\times$10$^8$ solar mass \citep{Melo2023}.
This lensing system is among those continuously monitored at optical wavelengths within the the COSmological MOnitoring of GRAvItational Lenses (COSMOGRAIL) project \citep{Courbin2005, Eigenbrod2006, Bonvin2016, Courbin2011, Awad2023}. This lensed AGN shows both short-term and long-term variability in the total optical flux (20\% over ~2 months and 1 magnitude over $\sim$12 years, respectively), making it a good system for estimating the Hubble parameter ($H_0$) from gravitational time delays. Indeed, it is
one of the five strong lenses used to estimate $H_0$ by the H$0$LiCOW collaboration ($H_0$ Lenses in COSMOGRAIL’s Wellspring, \citealt{Suyu2017}).

HE0435 has been among the known lensing systems reported in the first data release (DR1) of the \textsl{Gaia} mission  \citep{GaiaDR1,Ducourant2018a}, with only two lensed images detected (images A and C, ref Fig.\ref{fig:chandra_hst}). The second data release (DR2, \citealt{GaiaDR2}) reported the detection of all four images. \citet{Ducourant2018} used the sub-milliarcsecond (mas) astrometric precision \citep{Lindegren2021} of the lensed images positions to determine the lens mass model parameters. Such high angular resolution observations significantly improved on the previous lens model based on \textsl{Hubble Space Telescope} (HST) observations (e.g., Fig. 4 of \citealt{Ducourant2018}).

This lensed AGN was observed at X-rays with \axaf\ for the first time in 2006 \citep{Pooley2012}. Since then, ten more X-ray observations have been performed and used for microlensing studies in this system \citep{Pooley2012,Chen2012}.

\vspace{0.2cm}
In this article we use $H_0 = 67.8\ \mathrm{km\ s^{-1}\ Mpc^{-1}}$, $\Omega_M = 0.308$, and $\Omega_\Lambda = 0.692$  \citep{Planck2016}.


\begin{table}
\begin{center}
\caption{\textit{Chandra} X-ray  observations of HE0435 analyzed in this work. \label{table:chandra_observations}}
\begin{tabular}{|c|r|r|c|}
    \hline
    Obs ID&Live time (ks) & Counts\tablenotemark{a} & Start Date \\
    \hline
    7761&9.99 & 526 & 2007-12-18\\
    11550&12.88 & 452 & 2009-12-07\\
    11551&12.76 & 383 & 2010-07-04\\
    11552&12.76 & 258 & 2010-10-29\\
    14489&37.17 & 1382 & 2012-11-28\\
    14490&36.19 & 1261 & 2013-03-31\\
    14491&36.19 & 1262 & 2013-08-14\\
    14492&35.53 & 1068 & 2013-09-21\\
    14493&36.22 & 1225 & 2014-03-07\\
    14494&36.19 & 1135 & 2014-04-10\\
    23798&19.27 & 776 & 2021-04-02\\
    \hline
    Sum&285.15&9728&  \\
    \hline
\end{tabular}
\end{center}
\tablenotetext{a}{Total X-ray counts in a 6\arcsec$\times$6\arcsec box containing the four images. Includes an expected background count of 0.66 to 1.90.}
\end{table}

\section{Observations}

In this section we present additional multi-epoch \axaf\ X-ray observations and optical observations of HE0435 from the third \textsl{Gaia}  data release (DR3, \citealt{Gaia2016,Gaia2023}) that we used for our analysis. 

\subsection{X-ray Chandra observations}\label{sec:xray_observations}

\begin{figure*}
    \centering
    \includegraphics[width=\textwidth]{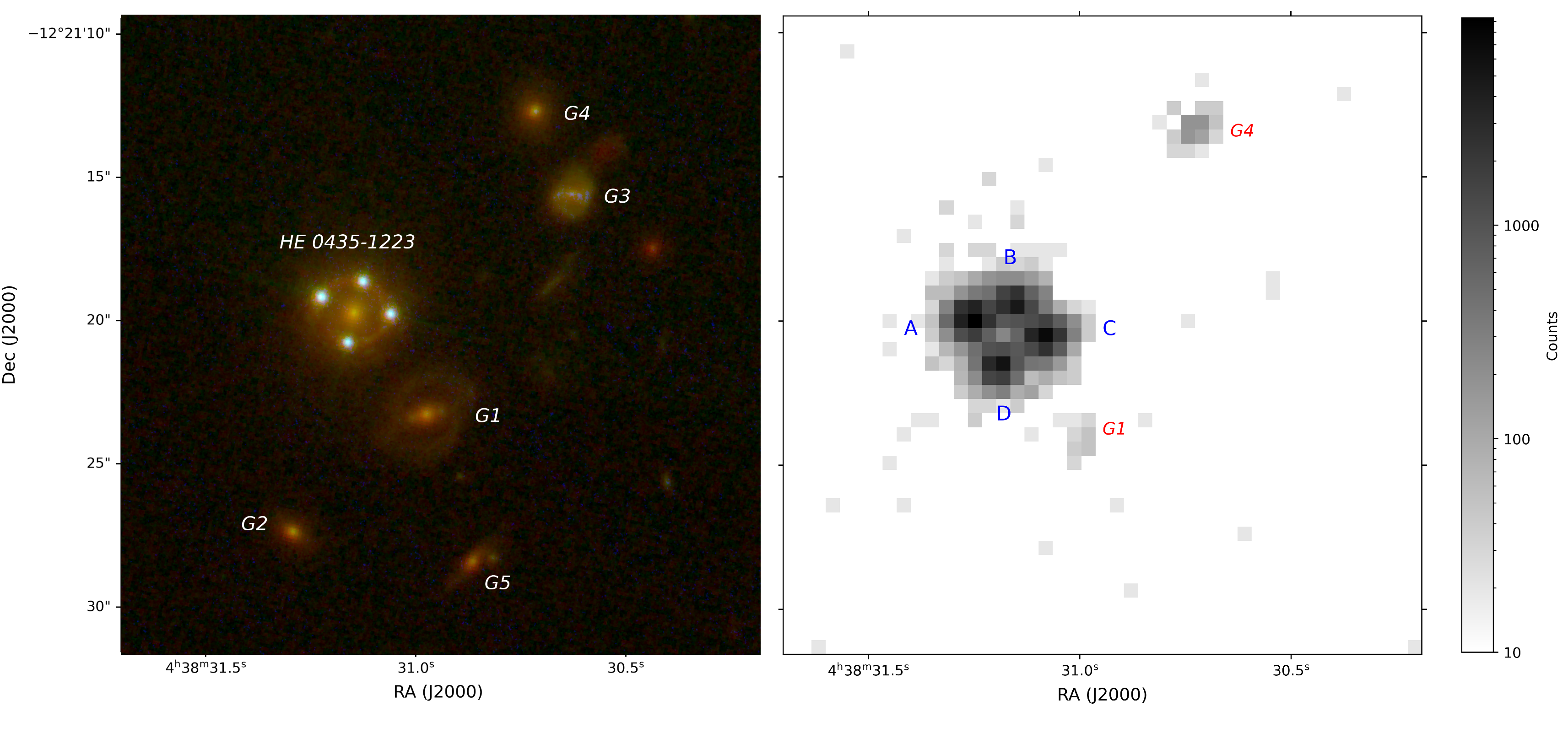}
      \caption{\textsl{Left}:  RGB image using \textsl{HST} NICMOS F160W (red), WFPC2 F814W (green) and WFPC3 F275W (blue) filters. We label the galaxies that \citet{Wong2017} identified as most relevant perturbers to HE 0435-1223. The cutout is 23\arcsec$\times$23\arcsec.  North is up, East is left. \textsl{Right:}  \axaf\, 0.5 -- 7.0 keV image of HE 0435-1223 field, binned into 0\farcs492 pixels. This is the merged image from the eleven observations in Table 1. The cutout is 23\arcsec$\times$23\arcsec, covering the same area of the RGB image on the left. North is up, East is left.  G1 and G4 are the only two field galaxies detected at X-rays and their properties are reported in Sec~\ref{sec:xray_observations}. The merged image is not used for any of the astrometric determinations in this paper.
    }
    \label{fig:chandra_hst}
\end{figure*}

HE0435 has been extensively observed at X-rays with \axaf\, and Table~\ref{table:chandra_observations} lists all of the observations  of this source which are available at \dataset[DOI: 10.25574/cdc.150]{https://doi.org/10.25574/cdc.150}.
All observations were performed in Very Faint mode, but we analyze only the 3 x 3 pixel data that would be telemetered in Faint Mode. ObsID's 7761 and 23798 were done with full frame readout, with 3.0 (one chip) and 3.1 (three chip) second exposures, respectively. The other ObsID's were done with 5 chips and 1/2 sub-chip readout, giving 1.70 second exposures. 

The data is from reprocessing run 5, and we use CIAO 4.13 \citep{Fruscione2006} and SAOImageDS9 version 8.4.1 \citep{Joye2003} for X-ray analysis. For the maximum likelihood astrometric analysis described below, we generated four simulated X-ray images at the positions specified by the \emph{Gaia} source positions in 
Table~\ref{table:lensmodel_params}. We use SAOTrace-2.0.4\_03 \citep{Jerius2004} to give a high fidelity distribution of X-rays exiting the mirror. SAOTrace is built upon extensive manufacturing metrology \citep[][]{vanSpeybroeck1997}, prelaunch X-ray measurments in the NASA/MSFC X-ray Calibration Facility \citep{Weisskopf1997,Odell1998}, and on-orbit observations \cite[][cf. figures 3, 10, 11, 12]{Jerius2000}. Section 3 of \citet{Jerius2000} explains how these calibration measurements are incorporated into SAOTrace. The model uses the synchrotron measurements of the energy-dependent reflectivity \citep{Graessle1998,Graessle2004}, properly weighting the contribution of each of the four shells. SAOTrace\footnote{https://cxc.cfa.harvard.edu/cal/Hrma/SAOTrace.html} is maintained and provided to the astronomical community by the Chandra X-ray Center. That simulation uses the measured spectrum and flux of each image, which we fit to a power law with fixed galactic absorption of $n_H =5.05 \times 10^{20} cm^{-2}$ \citep{Dickey1990}. The actual aspect solution is applied, incorporating the dither to give a distribution of counts in the focal plane. We use Marx-5.5.2 \citep{Davis2012} with the energy dependent sub-pixel event redistribution (EDSER) option, and adding an aspect blur of $0\farcs288$.

We show the merged wide-field X-ray image in Fig. \ref{fig:chandra_hst}. 
This was generated following the \axaf\ thread\footnote{https://cxc.cfa.harvard.edu/ciao/threads/combine/} for merging observations, and using the scripts \textsl{reproject\_events, dmmerge}. Because offsets in the different observations generate additional image blurring, we do not use this merged image in any of the astrometric analysis that follows.
We detect all of the four lensed images at high significance, and we report for the first time the X-ray detection of the nearby galaxies G1 ($z=0.7821$) and G4 ($z=0.4568$), following the nomenclature of \citealt{Wong2017}.  We measure a flux of $f_{\rm G1} = 1.5 \times 10^{-15} $ erg cm$^{-2}$ s$^{-1}$  and $f_{\rm G4} = 3.1 \times 10^{-15}$ erg cm$^{-2}$ s$^{-1}$, for G1 and G4 respectively. The corresponding rest frame 0.5--7 keV luminosities are $L_{X,\,G1} = 4.5 \times 10^{42}$ erg s$^{-1}$ and $L_{X,\,G4} = 2.5 \times 10^{42}$ erg s$^{-1}$.

\subsection{Optical Gaia DR3 observations}

All of the lensed images of HE0435 have been detected in the \textsl{Gaia} DR2, but \textsl{Gaia} DR3 shows significant improvements with respect to the previous detection. The new \textsl{Gaia} data release improves on the limiting sensitivity as faint as magnitude 21, providing better astrometry, photometry and list of radial velocities from Gaia DR2. There is also an improvement on the parallax precisions by 30\%, and on the proper motion precisions by a factor of 2. The details of the astrometric processing for the \textsl{Gaia} DR3 are described in \citet{Lindegren2021}.

For our strong lensing analysis, the most important improvement resulting from the DR3 data is the significantly better astrometric precision on the lensed images positions, which are the main constraints to the mass density distribution of the lens (Sec. \ref{sec:lensmodel}).

The \textsl{Gaia} DR3\footnote{\url{https://gea.esac.esa.int/archive/}} position and uncertainties on the four lensed images of HE0435 are listed in Table \ref{table:observed_and_model_positions}.

\section{Lens Modelling}
\label{sec:lensmodel}

\citet{Ducourant2018} demonstrated how the mass model for HE0435 could be improved by using the \emph{Gaia} DR2 positions as compared to the lower angular resolution \textsl{HST} data. Here, we further refine the lens mass model model using  the \emph{Gaia} DR3 \citep{Gaia2016,Gaia2023}
for the four image positions. Columns 2 and 3 of Table~\ref{table:observed_and_model_positions} give the \emph{Gaia} DR3 positions relative to image A.  We use the public software \textit{gravlens}  \citep{Keeton2001} to model the \emph{Gaia} DR3 lensed images positions, without fitting for their relative flux ratios, as they could be strongly affected by microlensing by stars in the lensing galaxy (see Sec. \ref{sec:discussion}).

We adopt a power-law elliptical mass density distribution with an additional external shear component, starting from the best available model of \citet{Ducourant2018}. We keep the slope ($\alpha$) of the power-law fixed at 1, hence assuming an isothermal model, to match the model of \citet{Ducourant2018}.
Moreover, we fix the external shear strength, and the position angles of the external shear and ellipticity at the values of \citet{Ducourant2018}.
As a first iteration we allow the Einstein radius, and the ellipticity to vary, and minimize $\chi^2$. Then we hold those parameters fixed, and further minimize $\chi^2$ by allowing the position of the ellipsoid to vary.

\begin{figure}
    \centering
    \includegraphics[width = 0.45\textwidth]{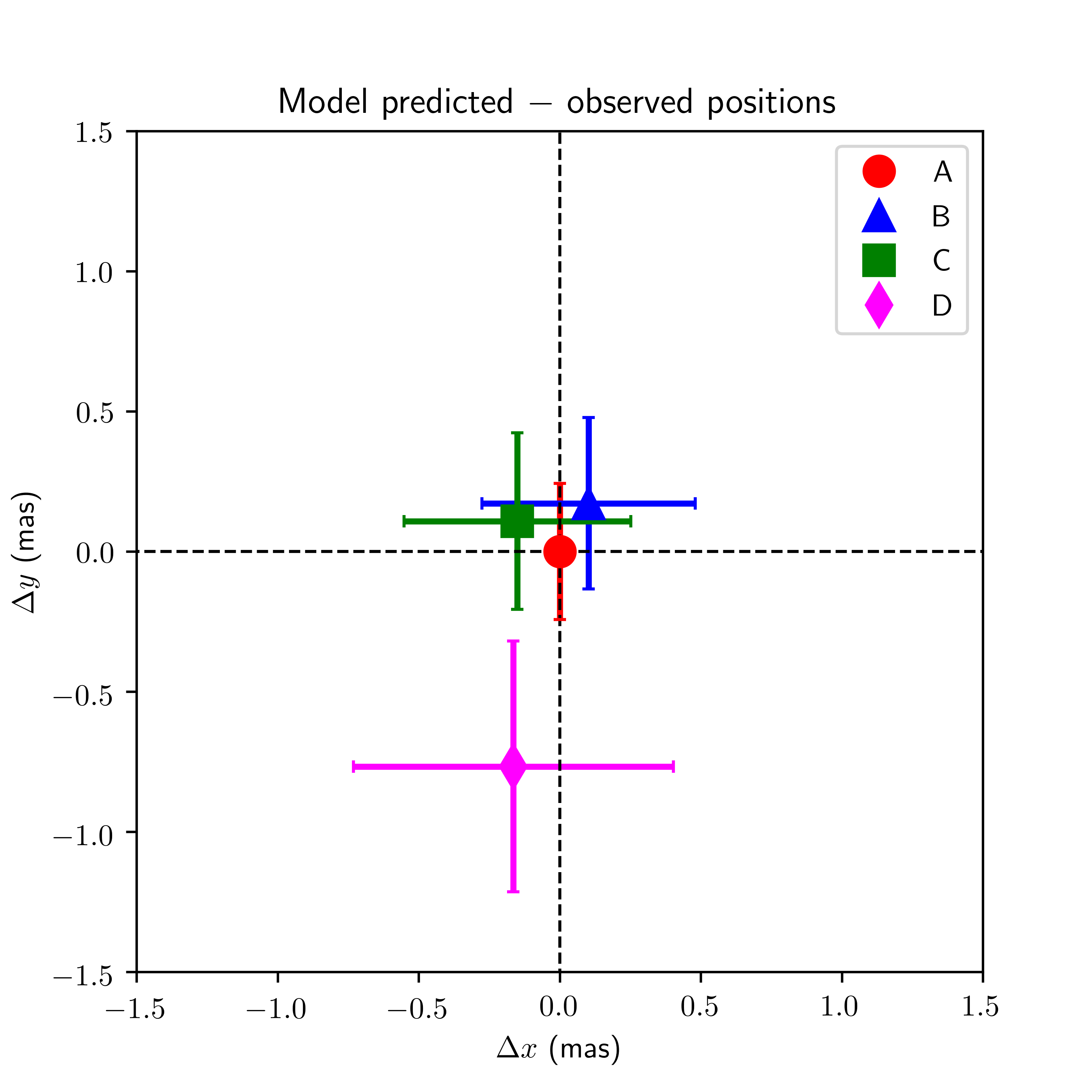}
    \caption{Offsets between the model-predicted and observed positions in units of mas for each image (as indicated in the legend shown in the top right). The error bars are the 3$\sigma$ astrometric uncertainties on the lensed images position as measured by \textsl{Gaia} DR3 and reported in Table \ref{table:observed_and_model_positions}.}
    \label{fig:images_offset}
\end{figure}

We infer the best-fitting values and the uncertainty on the lens model parameters that we optimized for (68 per cent confidence level) from a Markov-Chain Monte Carlo (MCMC) sampler implemented in \textit{gravlens}. The results of our lens model are given in Table \ref{table:lensmodel_params}, and the difference between the observed
and model-predicted image positions is shown in Fig. \ref{fig:images_offset} and listed in Table \ref{table:observed_and_model_positions}.

Table~\ref{table:lensmodel_params} also reports the best model parameters from the \emph{Gaia} DR2 \citep{Ducourant2018}. This lens mass model can reproduce all of the lensed images except image D within their $3\sigma$ astrometric uncertainties (see Fig. \ref{fig:images_offset} and Table \ref{table:observed_and_model_positions}. We highlight that the primary goal of this work is to investigate the relative location of the X-ray and the optical emission, and that the same mass model is used for estimating the source position and uncertainty at optical and X-ray wavelengths. Therefore, this lack of the model is a systematic applied to both optical and X-ray source reconstructions, and does not affect our study on the relative location of the two emitting regions (see also Sec.~\ref{sec:discussion}).

For the fixed mass model that we adopt, we calculate the uncertainty in the source plane position of the Gaia emission. At each of a grid of positions with 0.04 mas spacing, we compute chi-squared for the deviation of the image positions predicted by \textit{lensmodel}, from the measured Gaia positions. We compute the contours of chi-squared probability corresponding to 1, 2, or 3 $\sigma$ of a Gaussian distribution. This gives a 1$\sigma$ source location error of approximately 0.1 mas in each coordinate. Note that the uncertainties of the mass model parameters will cause the Gaia source position to change relative to ICRS coordinates, but those uncertainties will not change the relative location of the X-ray to optical emission. 

\begin{table*}
    \begin{center}
    \begin{tabular}{|c|c|c|c|c|c|}
    \hline
     & \multicolumn{2}{c}{GAIA DR3  Positions}&\multicolumn{3}{|c|}{Model-predicted Positions}\\
     \hline
     Lensed image    &RA offset  (mas) & Dec offset (mas)  & RA offset (mas)  & Dec offset (mas) & Difference (mas)\\
     \hline
    A    & $0.000\pm0.098$ & $0.000\pm0.081$ & $-0.1181$ &-0.0819  &0.144 \\
     B    & -1476.56$\pm$0.126 & 552.94$\pm$0.102 & -1476.6730 &553.1614  &0.249 \\
       C   & -2466.27$\pm$0.134 &-603.05$\pm$0.105  &-2466.2060  &-603.3546  & 0.311\\
      D   & -938.66$\pm$0.189 &-1614.43$\pm$0.149  & -938.4350&-1615.2520&0.852 \\
      \hline
    \end{tabular}
    \caption{{Column 1: ID of the lensed image; Columns 2 and 3: \textsl{Gaia} DR3 positions in RA \citep{Gaia2016,Gaia2023} and Dec with their $1\sigma$}  uncertainties, relative to image A (mas). The \emph{Gaia} Archive lists the position of image A at Right Ascension (RA)= 69.56230463 deg, Declination (Dec) = $-$12.28731415 deg. Columns 4 and 5: lens model-predicted positions relative to the \emph{Gaia} position of image A. Column 6:  distance between the \emph{Gaia} DR3 image position and the relative model-predicted lensed images. All offsets are arc length on the celestial sphere in units of mas and also shown in Fig. \ref{fig:images_offset}.} \label{table:observed_and_model_positions}
          \end{center}
\end{table*}

\begin{table*}
\begin{center}
\begin{tabular}{ |l|c|c|c| } 
 \hline
\textsl{Parameter} & \textsl{Symbol} & \textsl{Gaia DR2}&\textsl{ Gaia DR3 (this work)}\\
 \hline
    Slope & $\alpha$ & $\equiv 1$ & $\equiv 1$ \\
    Mass strength (arcsec)    & $b$ &  $1.2\pm0.002$ & $1.19778\pm0.0002$\\ 
    Galaxy RA from imageA (arcsec) &$\Delta x_L$ &  $-1.1766\pm0.0003$ & $-1.17655\pm0.0002$\\ 
    Galaxy Dec from imageA (arcsec) & $\Delta y_L$ &  $-0.5536\pm0.0012$ & $-0.55342\pm0.0004$\\
    Ellipticity &$e$ &  $0.079\pm0.01$ & $0.0801\pm0.002$\\
    Angle E of N (degrees) & $\theta_e$ &  $290.5\pm0.8$ & $\equiv290.5$\\
    Shear&$\gamma$ &  $0.088\pm0.002$ & $\equiv 0.088$\\
    Angle E of N (degrees) &$\theta_\gamma$ &  $15.9\pm0.1$ & $\equiv 15.9$\\
    \hline
    Source position RA from image A (arcsec) & $\Delta x_s$ & $-1.1817 \pm$0.0001 & $-1.181640 \pm$0.0001 \\
    Source position Dec from image A (arcsec) &  $\Delta y_s$ & $-0.5216 \pm$0.0001 &$-0.5218143 \pm$0.0001 \\
    \hline 

    \hline
\end{tabular}
\caption{ Minimum-$\chi^2$ parameters of the parametric lens model for HE0435 reported in \citealt{Ducourant2018} (\emph{Gaia} DR2) and this paper (\emph{Gaia} DR3). We adopted a power-law mass density model, where $b$ is the lens strength (arcsec); $\Delta x_L$ and
 $\Delta y_L$ are the positions in RA and Dec of the lensing galaxy relative to image A (arcsec); $e$ is the ellipticity,; $\theta_e$ is the position angle of the ellipticity (east of north, degrees),  $\gamma$ is the external shear strength, and $\theta_{\gamma}$  is the external shear position angle (east of north, degrees). The density slope of the ellipsoidal power-law mass distribution is given by $\alpha$, which has been kept fixed to 1 to obtain an an isothermal profile as \citet{Ducourant2018}. $\Delta x_s$ and $\Delta y_s$ are the RA and Dec source coordinates relative to image A. 
}
\label{table:lensmodel_params}
\end{center}
\end{table*}

\section{X-ray Astrometric Analysis} \label{sec:xray_localization_method}

We apply a double maximum likelihood process, based on  the Bayesian lensing analysis described in \citet{Schwartz2021} and \citet{Spingola2022}.  The algorithm is described in detail in Appendix A. We do not attempt to make direct measurements of the positions of the individual four images. Instead we make a model predicting the expected number of counts per pixel in each image. The model has two free parameters, the (RA, DEC) of the X-ray quasar in the source plane. The lens mass model that correctly produces the positions of the four Gaia images is used to determine where the X-ray images must be located as a function of the quasar source plane position.

For each ObsID we consider a 6\arcsec $\times$ 6\arcsec region. This contains the four images, with about a 1\arcsec\  margin outside the FWHM response. We sort the data into one half the native ACIS pixels, namely 0\farcs246 $\times$ 0\farcs246 bins. We sort data from the four simulated images into the same bin structure, with each image having an independent free normalization constant and also adding the expected background per bin as determined from a much larger region containing no recognized sources. We adjust the normalization of each simulated image to maximize the probability $P$ of observing the actual counts in each of the $24 \times 24 = 576$ pixels, and compute the maximum likelihood as 
\begin{equation}
C=-2P =-2 \sum_{i=1}^{576}\ln(\lambda_i^{n_i} \exp(-\lambda_i)/n_i!).
\end{equation}
The $\lambda_i$ are the sum of the contributions from each of the four simulated images with their best fit normalization parameter, and with the total predicted counts constrained to equal the total observed number. Note that the four amplitudes are uninteresting parameters in the statistical sense. 

For the first maximum likelihood test we correct for the inaccuracy of the \axaf\ absolute celestial location by rastering the predicted counts relative to the observed, readjusting the normalizations at each step, and adopt the aspect correction corresponding to the minimum in equation 1. We compute the location uncertainty using the theorem of Wilks \citep{Wilks1938,Cash1979}, that the differences of the likelihood function in equation (1) are distributed as $\chi^2$ with two degrees of freedom, for the two interesting parameters of right ascension and declination aspect solution offset. 

\begin{figure*}[t]
    \centering
    \includegraphics[width = \textwidth]{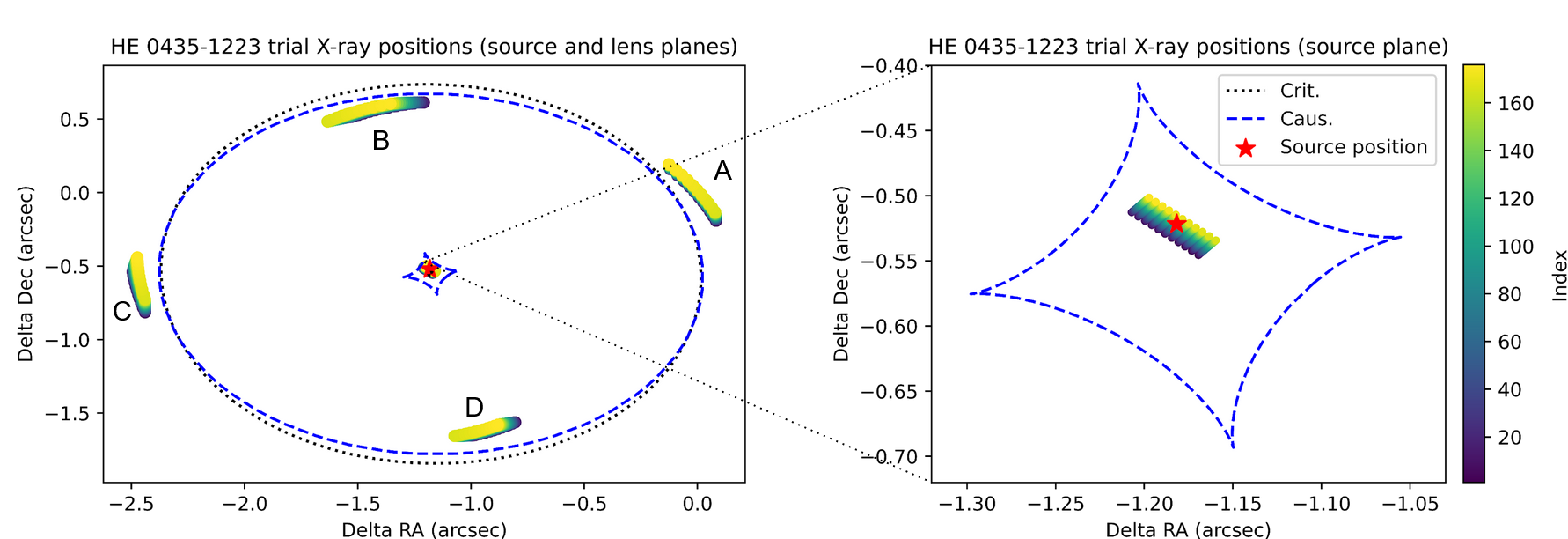}
    \caption{ \textsl{Left:} Lens mass model for HE0435.  The lens critical curve is shown by the black dotted line, while the source plane caustics are indicated by the dashed blue lines. The red star indicates the best optical source position (listed in Table \ref{table:lensmodel_params}).The colored points correspond to the simulated lensed images (A, B, C and D) corresponding to each of the alternative X-ray source positions. \textsl{Right:} Source plane reconstruction for HE0435. The dashed blue line indicates the caustic curve and the red star shows the best optical position given our best lens model. 
    The colored points consist of an 11 by 16 array of putative X-ray positions, quasi-perpendicular and parallel to the upper left (NW) section of the caustic. We use the lens model to map from these trial source positions to the image positions in the left panel.    
    The color bar on the right codes these 176 trial positions. North is up and East is right in both figures.} \label{fig:CritPlot}
\end{figure*}

For the second maximum likelihood test we consider an array of putative X-ray source positions surrounding the position deduced from the \emph{Gaia} astrometry. Figure~\ref{fig:CritPlot}, right, indicates the array of 11 positions spaced 5 mas apart perpendicular to the caustic, and 16 positions spaced 1 mas apart parallel to the caustic. Figure~\ref{fig:CritPlot}, left, shows the mapping of each source position to the image plane. This mapping shows how the different possible source positions lead to different separations of the predicted image positions. At each position we adjust the $\lambda_i$ for the new relative image positions, carry out the first maximum likelihood to best register the \axaf\ aspect solution, and compute the best maximum likelihood value for the case that the trial source position is the true X-ray source position. For each ObsID, we find the minimum of equation (1) and use third order interpolation to find the contours of confidence $p$ that give the $\chi^2$ probability of equaling $p$ for two degrees of freedom. 

\section{Results}\label{sec:results}

\subsection{Relative position of optical and X-ray emissions }\label{sec:X-ray_location}

\begin{figure*}[t]
    \centering
    \includegraphics[width = \linewidth]{
    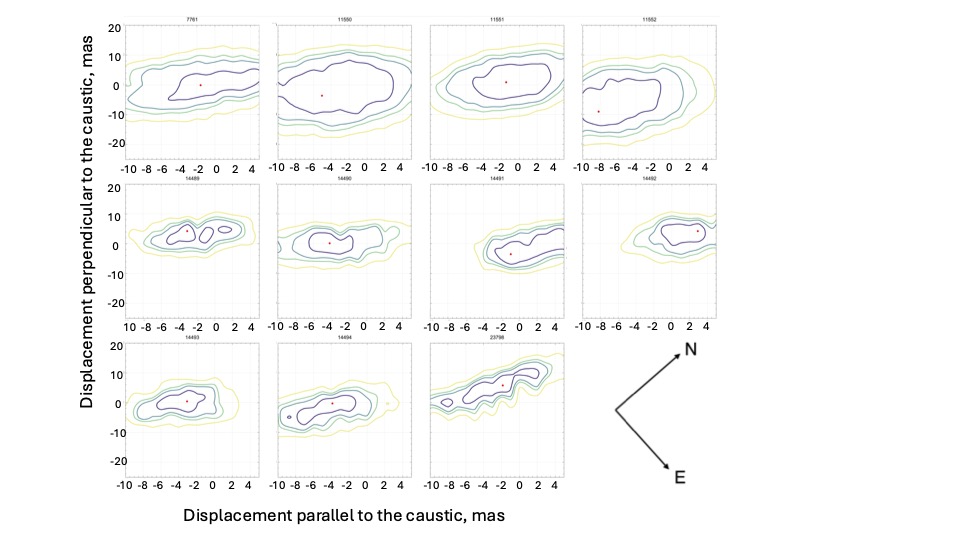}
    \caption{Position contours derived from each of the 11 separate ObsID's, in the respective order given in Table~\ref{table:chandra_observations}. The final panel shows the orientation relative to celestial coordinates, rotated by 50.04\degs\ east of north about the (0,0) coordinates of the \emph{Gaia} source position. Contours include the X-ray emission centroid to a confidence of 68\%, 90\%, 95\%, and 99\%, respectively, moving outwards, based on the $\chi^2$ probability of the contour value exceeding the minimum value at the red "x". }
        \label{fig:positions}
\end{figure*}

Figure~\ref{fig:positions} 
shows the confidence contours for the location of the HE0435 source at each X-ray \axaf\ observation (listed in Table \ref{table:chandra_observations}).
The contours indicate a third order interpolation from the $11\times16$ grid of trial X-ray source locations. They mark the $\chi^2$ probability for deviating by more than 68\%, 90\%, 95\%, and 99\%, respectively, from the minimum value. We have indicated the maximum likelihood within that grid as the best location of the X-ray emission for each observation. Note the presence of alternative relative minima in some of the observations. The rectangular coordinate system is parallel and perpendicular to the caustic nearest the source position deduced by \emph{Gaia} DR3, and at an angle of 50.04\degs\ east of north. That orientation is also shown in the bottom right panel.   

\begin{figure*}
    \centering
    \includegraphics[width = 0.6\textwidth]{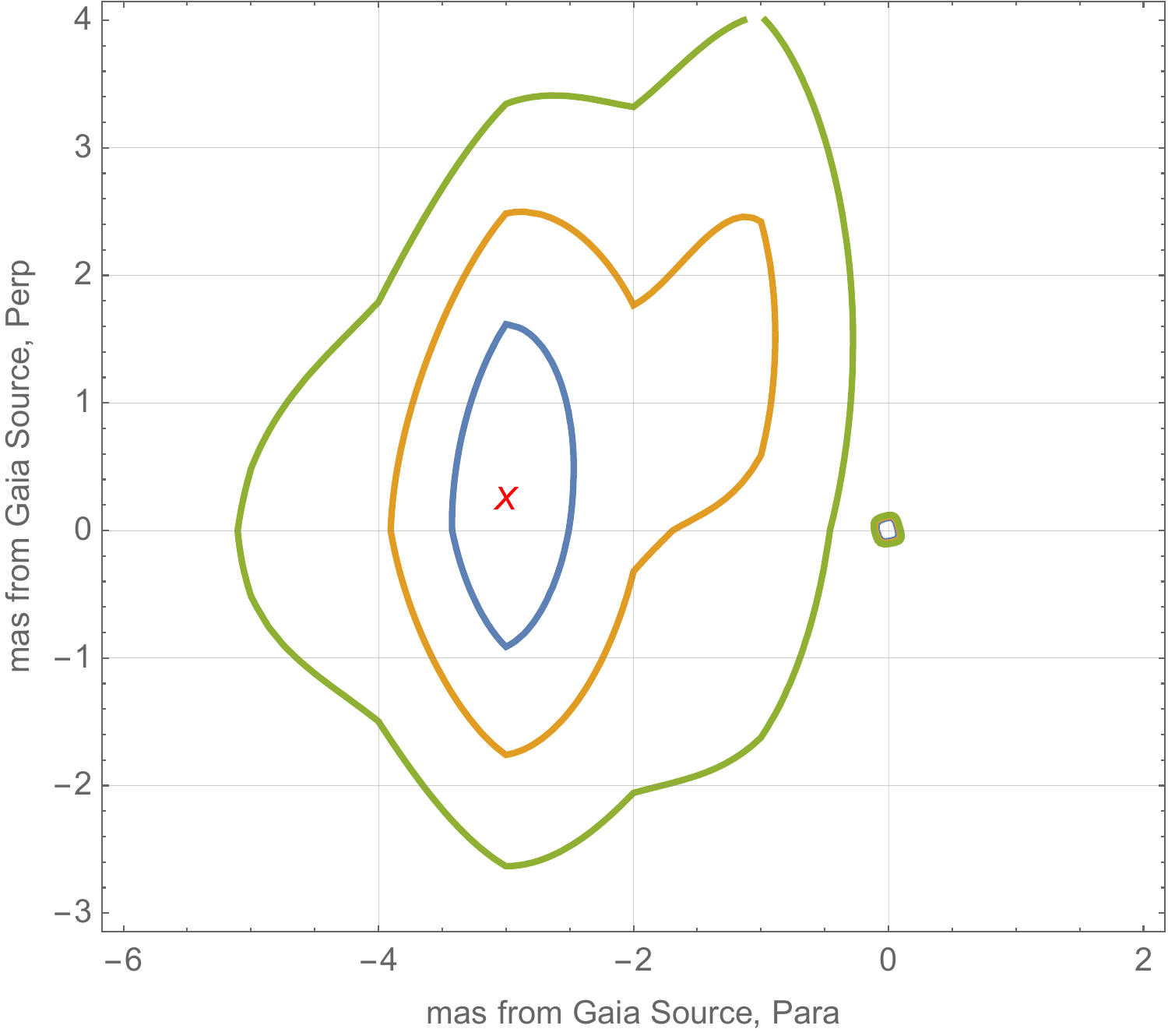}
    \caption{Likelihood contours for the X-ray source position (red ``X") from all observations, assuming it has remained fixed. Contours include the X-ray emission centroid to a confidence equivalent to 1, 2, 3 $\sigma$, respectively, moving outwards. The blue, red and green contours at (0,0) show the 1, 2, 3 $\sigma$ of the deduced source position from the positions of the \emph{Gaia} images. Celestial coordinate orientation is the same as shown in the final panel of figure \ref{fig:positions}, namely at an angle of 50.04\degs\ east of north with respect to Fig.~\ref{fig:CritPlot}.}
        \label{fig:allpositions}
\end{figure*}

Assuming a single, fixed X-ray emitting region throughout the 13.3 years, Fig.~\ref{fig:allpositions} shows the maximum likelihood probability contours and best X-ray location summed for all 11 \axaf\ observations. To better than 3$\sigma$ statistical confidence, the X-ray location is displaced 3 mas from the deduced optical position. We will discuss the confidence of this separation, and possible implications of the disjoint positions in Section~\ref{sec:discussion}. Table~\ref{table:positions} tabulates the maximum likelihood center position from each ObsID, the probability that the position is consistent with the best maximum likelihood X-ray position deduced from the sum of all 11 observations, and the probability that it is consistent with the position of the \emph{Gaia} DR3 source. More precisely, these are the probabilities that the source is found at the measured position or further, if the true position is as hypothesized.  
The statistical probability that the sum of all observations is consistent with the \emph{Gaia} position is contradicted at the equivalent of a  3.36 $\sigma$ deviation of a Gaussian distribution. This implies some X-ray emission region(s) to be disjoint from the quasar core, while still allowing substantial X-ray emission from the quasar. The offset between the X-ray and the optical emitting regions is of $3.0\pm0.5$ mas (1$\sigma$), which at the redshift of the source corresponds to $\sim26$ pc.

\begin{table*}
\begin{center}
\caption{Offsets and probabilities of the X-ray positions \label{table:positions}} 
\begin{tabular}{|c|r|r|c|c|}
    \hline
    Obs ID & $\Delta$RA\tablenotemark{a} (mas)&$\Delta$Dec\tablenotemark{a} (mas)& P\tablenotemark{b} (Best Position) & P\tablenotemark{b} (\emph{Gaia} Position) \\
    \hline
    7761&-0.99 & -1.18 & 0.573 & 0.569\\
 
    11550 &-0.63 & 0.37 & 0.994 & 0.881\\
    11551 &-1.27 & -0.34 & 0.638 & 0.653\\
    11552 &0.47 & -6.91 & 0.556 & 0.218\\
    14489&-2.71 & 0.96 & 0.345 & 0.129\\
    14490&-2.66 & -3.17 & 0.525 & 0.143\\
    14491 &2.15 & -3.10 & 0.159 & 0.305\\
    14492&-1.50 & 5.12 & 0.067 & 0.353\\
    14493&-2.43 & -2.01 & 0.971 & 0.087\\
    14494&-1.65 & -1.96 & 0.936 & 0.143\\
    23798&-6.52 & -0.74 & 0.167 & 0.004\\
    \hline
   All&-2.99&0.16  & 1& 0.00039\\
    \hline
\end{tabular}
\end{center}
 \tablenotetext{a}{Offsets from the \emph{Gaia} source position RA,Dec(J2000)=4$^h$38$^m$14\fsecs8725, -12$^d$17$^m$14\fsecs8528, where the measured source is at \emph{Gaia} position +($\Delta$RA,$\Delta$Dec).}
 \tablenotetext{b}{Probability that the maximum likelihood source position from this ObsID is consistent with the indicated position.}
\end{table*}

\subsection{X-ray Varstrometry}\label{sec:xray_varstrometry}

The astrometric jitter caused by optical variability of AGN has been used to deduce the presence of offset and multiple sources, even without being able to resolve them individually \citep{Shen2019, Hwang2020, Shen2021, Chen2022}. The term \textsl{varstrometry} was coined by \citet{Hwang2019} to mean the use of variability and astrometry, citing previous use of a similar technique to study binary stars \citep{Wielen1996}, and suggesting that the technique could be applied to other wavebands and sources. When at least one of two sources separated by an angular distance, $\theta$, is variable in multiple observations, their unresolved but precisely measured centroid will change when observed at different times. In a simple situation of two sources with mean flux ratio $q\le1$, mean total flux $<F>$ and random, independent flux variance $\sigma^2_F$, equation 5 of \citet{Hwang2020} gives the root-mean-square (rms) jitter in the astrometric position to first order as

\begin{equation}\label{varEQ}
 \sigma_{astro} =\theta \frac{ q}{1+q}\sqrt{\sigma^{2}_{F}}/<F>. 
\end{equation}

This is easily generalized to somewhat more complicated cases \citep{Hwang2020}. 
This technique has been successfully applied to \emph{Gaia} data to find dual quasars based on the sub-milli-arcsec (mas) precision of the optical centroids (e.g., \citealt{Chen2022}).
In this work we apply this technique to search for an evidence for a dual AGN in the source plane of HE0435.

We model two sources of equal mean flux causing the positions measured for HE0435 to vary throughout the years. In this picture one source is fixed at the \emph{Gaia} position, and our X-ray position is measuring the centroid of the two AGN. Random variability of the two sources will lead to an rms jitter, $\sigma_{astro}$, along the line of the two sources, and $\sigma_{astro}$ will be signficantly larger than position jitter, $\sigma_{noise}$ in the orthogonal direction. 
From Table~\ref{table:positions} we can calculate $\sigma_x=2.4$ mas along the line from the \emph{Gaia} position to the best centroid, and jitter perpendicular is $\sigma_y=3.3$ mas from the maximum likelihood position of all the data. We take a 1$\sigma$ upper limit $\sigma_{astro}=3.3$ mas. 

\begin{figure*}[t]
    \centering
    \includegraphics[width = 0.8\textwidth]{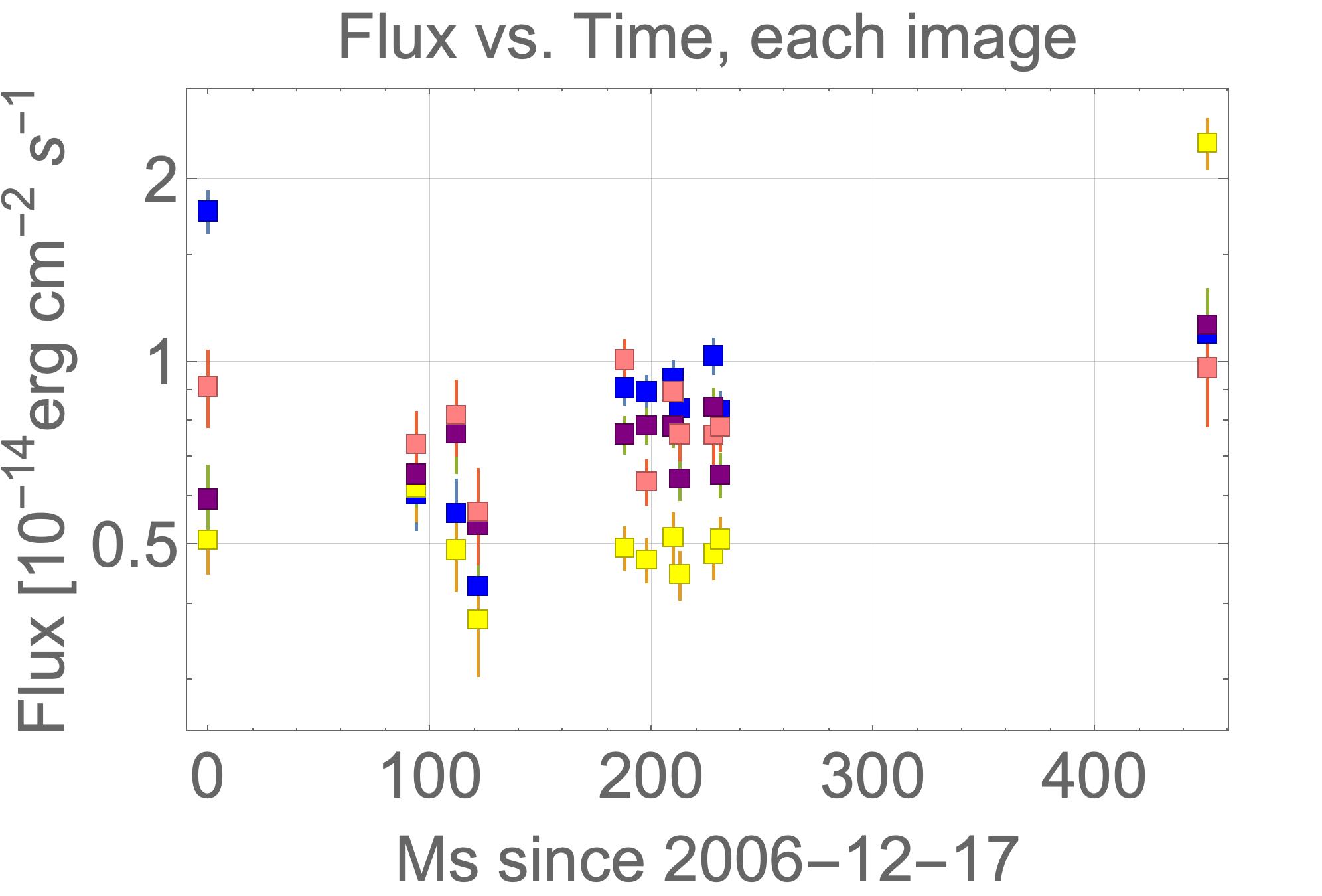}
    \caption{Measured 0.5 to 7 KeV flux of each image of HE0435 vs time since the initial observation. Fluxes are corrected for the predicted magnification at the best image position. Image A, B, C, and D fluxes are blue, yellow, purple, and pink, respectively.  Error bars are derived from the square root of the number of counts. The quasar shows variability on months/years time scales. We attribute the first image A point and the last image B point to microlensing.} \label{fig:fluxVtime}
\end{figure*}

Table~\ref{table:fluxes} gives the demagnified, unabsorbed 0.5 -- 7 keV flux for each image, for each ObsID. The fluxes are derived from power law fits of counts in an 0.9 arcmin radius about the centroid of each image. Images A, B, C, and D are demagnified according to the magnifications predicted by our model, 9.1, 9.8, 9.2 and 6.3, respectively. Poisson statistical errors vary from 5\% to 13\%  for the individual measurements. Image ratios involving image A in ObsID 7761 and image B in ObsID 23798 are discordant by more than a factor of two from expected values of 1 to 1.5. This is apparent in Fig.~\ref{fig:fluxVtime} that shows the  measured fluxes of each image as a function of time. 
We attribute those to microlensing, and ignore those two ObsID's to report the average flux and flux variability (see Sec. \ref{sec:xray_variability_discussion}). Fig. \ref{fig:fluxVtime} also shows intrinsic variations of the quasar. 

Considering ObsIDs 11550 through 14494, the rms flux variation is 14\%.
The average of the summed, demagnified flux of the four images in those ObsIDs is $2.75\pm0.38\times10^{-14}$ erg cm$^{-2}$ s$^{-1}$ which corresponds to a luminosity of $5.52\pm0.77\times10^{44}$ erg s$^{-1}$ at the redshift of $z=1.689$. The quoted errors are the rms of the nine different fluxes from the ObsID's used. The error on the overall mean due to Poisson statistics would be 1\%. 

We can now use equation~\ref{varEQ} with $q=1$ for equal mean fluxes and conclude that two such quasars are within 47 mas, or within 410~pc ($1\sigma$) at the distance of HE0435.  For highly variable quasar pairs and with a definitive measure of $\sigma_{astro}$ we see the potential to measure dual AGN at separations less than one kpc. In this context, we note that X-ray variability fractions are generally greater than those in optical or longer wavelength bands.

Although we report a null result here, this stands as the first attempt at X-ray varstrometry. It illustrates the potential for future investigations, even for systems that are not gravitationally lensed. 

\begin{table*}
\begin{center}
\caption{Demagnified flux for each image in each ObsID. \label{table:fluxes}} 
\begin{tabular}{|l|r|r|r|r|r|}
    \hline
   & \multicolumn{4}{|c|}{Demagnified flux of each image\tablenotemark{a}} &Source\\
ObsID	&	 A	&	B	&	C	&	D	&	flux\tablenotemark{b}	\\
    \hline
7761	&	17.67	&	5.08	&	5.92	&	9.09	&	26.8\\
11550	&	6.03	&	6.18	&	6.52	&	7.30	&	26.0\\
11551	&	5.61	&	4.89	&	7.62	&	8.14	&	26.2\\
11552	&	4.26	&	3.75	&	5.37	&	5.64	&	19.0\\
14489	&	9.03	&	4.93	&	7.57	&	10.04	&	31.6\\
14490	&	8.91	&	4.71	&	7.84	&	6.34	&	27.8\\
14491	&	9.40	&	5.13	&	7.81	&	8.90	&	31.2\\
14492	&	8.36	&	4.45	&	6.41	&	7.58	&	26.8	\\
14493	&	10.21	&	4.81	&	8.40	&	7.55	&	31.0\\
14494	&	8.29	&	5.09	&	6.50	&	7.80	&	27.7	\\
23798	&	11.09	&	22.91	&	11.49	&	9.75	&	43.1\\
\hline
Average\tablenotemark{c}&8.12	&	4.90	&	7.41	&	8.01	&	28.8\\
RMS\tablenotemark{c}&2.16	&	0.61	&	1.65	&	1.36	&	5.86\\
Percent Variability&27	&	12	&	22	&	17	&	20	\\
    \hline
\end{tabular}
\end{center}
\tablenotetext{a}{All fluxes in units of 10$^{-15}$ erg cm$^{-2}$ s$^{-1}$, observed in 0.5 -- 7 keV }
\tablenotetext{b}{The source flux is the sum of the fluxes of the four images corrected for their magnification factors. For the strongly microlensed fluxes of image A for ObsID 7761 and image B for ObsID 23789, the flux is replaced by the mean of the other three images.}
\tablenotetext{c}{Omitting ObsID 7761 for image A  and ObsID 23789 for image B  due to microlensing.}
\end{table*}

\section{Discussion}\label{sec:discussion}

\subsection{Lens models from sub-mas astrometric precision}\label{sec:precise_lens_model_discussion}

In Sec.~\ref{sec:lensmodel} we presented a refinement of the lens mass model from \citet{Ducourant2018} using the more precise positions of the lensed images of HE0435 
from the \textsl{Gaia} DR3. The position of the four lensed image are the only constraints to the lens mass model and are localised at an astrometric precision of 100s of $\mu$as (Table \ref{table:observed_and_model_positions}). The \textsl{Gaia} DR3 astrometric uncertainties on images A and D are a factor of two better than those from the DR2, while the uncertainties on the position of images B and C is about 10\% better than those reported from DR2 \citep{Ducourant2018}. This improvement in the astrometric precision strongly reduces the parameter space needed to be explored during the optimization of the lens model, inferring the set of parameters at sub-percent precision level. 
Using the \emph{Gaia} DR3 data, we improve on the lens model uncertainties up to an order of magnitude (for instance, in the case of the mass strength $b$ and $e$, see Table \ref{table:lensmodel_params}).

The precision of the lens model parameters obtained with the exquisite \emph{Gaia} DR3 astrometric uncertainties is comparable to that derived from  extremely high angular resolution radio observations with Very Long Baseline Interferometry (VLBI; e.g., \citealt{Biggs2004, McKean2007, Spingola2018, Hartley2019, Powell2021, Powell2022}). The main consequence of obtaining precise lens models in our study of the relative emission in sources at high redshift is described in the next section (Sec. \ref{sed:offset_discussion}).

We would like to highlight that despite the high statistical precision of our lens model, the systematic uncertainties due to the choice of the model (i.e., an elliptical power-law mass density profile) may be orders of magnitude larger. Nevertheless, this systematic uncertainty does not limit our aim of spatially differentiating the origin of multi-wavelength emission in high-$z$ AGN, as two intrinsically separated components will result in two spatially separated sources for any lens mass model. The amplitude of the separation is sensitive to the adopted mass density profile. But since any mass density profile must correctly predict the optical positions, the limit in spatially resolving the multi-wavelength emission is the angular resolution and sensitivity necessary to detect the offsets in the lensed images, rather than the accuracy of the chosen lens model (see \citealt{Barnacka2017} for an in-depth theoretical discussion on this application). 

As a check, we consider a simple elliptical potential from a Navarro-Frenk-White (NFW) model, nfwpot from \citet{Keeton2001} Table 3.3.  We found that a surface density parameter $\kappa_s = 0.95195$, ellipticity 0.0315 at 70.88\degs, and external shear 0.0835 at 15.63\degs reproduced the Gaia image positions within errors. This resulted in a deduced source plane position of the optical source that was about 1 mas from the position derived from the SIE model. Repeating the complete X-ray analysis using this model changed our deduced X-ray source position to (-2.55, -1) in the coordinates of Figure 5. The confidence contours of the two models overlap, and each 90\%  confidence contour includes the best X-ray position of the other model. The NFW model excludes the Gaia position at 3.76$\sigma$ confidence. We therefore conclude that the X-ray position robustly excludes the Gaia position, while allowing for some additional uncertainty to the confidence contours.

The singular isothermal elliptical (SIE) mass density distribution model has been developed to provide a more realistic representation of mass density profiles, taking into account both the baryonic and dark-matter distribution in lensing galaxies \citep[e.g.,][]{Kassiola1993, Kormann1994, Keeton1998},
which are typically early-type galaxies. Several works showed that lensing galaxies on average require a mass-density slope that is steeper than the NFW slope. For example, analysis of the SLACS lensing sample \citep{Bolton2008} revealed that these lenses are very well approximated by an SIE profile. Among the many works on this lensing sample, we highlight \citet{Koopmans2009} and \citet{Auger2010}, who concluded that the SIE is a good model for strong lenses using a sample of 73 objects (see also \citet{Sonnenfeld2013}).  Recently, \citet{Sonnenfeld2024} revised the analysis and demonstrated that this sample is well represented by an SIE profile even when carefully taking into account the selection effects of the survey.

A "steeper" mass density profile (like the one of the SIE) is to be expected as a natural consequence of the dissipative processes due to in situ star formation at the center of galaxies. This is even more enhanced in the case of lens galaxies that are parts of groups/clusters (see for instance \citet{Spingola2018} Sec. 4.3 for a detailed discussion on the topic, and also \citet{Auger2008} and \citet{McKean2010}), as in the case of HE0435 (Fig. 1 of this manuscript).
Moreover, the mass primarily responsible for the lensing effect ($\Sigma>\Sigma_{\rm crit}$) is the central part of the lenses, where the fraction of dark matter is small, favoring the assumption of an SIE rather than an NFW mass density profile.

\subsection{Optical-X-rays offsets on tens of parsecs}\label{sed:offset_discussion}

In Sec. \ref{sec:results} we showed evidence that the best-fit position of X-ray emission from HE0435 is $3.0\pm0.5$ mas from the optical position. Using the maximum likelihood ratio test this is equivalent to a $3.36\sigma$ deviation of a Gaussian distribution.  

Considering the possible 
systematics due to the choice of the lens mass model (see Sec. \ref{sec:precise_lens_model_discussion} for an in-depth discussion), we recognize this is indicative of an optical-X-ray offset at $\sim3$ mas. In any event, we note that a significant component of the X-ray emission is restricted to a size less than $\approx 10^{16}$ cm (i.e., one micro-arcsecond at the source distance) because of the observed microlensing and the intrinsic variability of the quasar emission.
At the redshift 1.689, the $3.0\pm0.5$ mas separation of the X-ray and optical positions would correspond to $26\pm4$ pc in projection. 
There are several reasonable explanations for such a separation, which include extended jets with bright knots, mass outflows and AGN pairs.

\vspace{0.15cm}

A plausible explanation of an X-ray to optical offset might be that the X-ray emission is dominated by a jet component, as was the case for the core of M87 for many years \citep{Harris2003,Harris2006,Harris2009}. 
Since M87 lies at the distance of only $\sim15$ Mpc it is possible to spatially resolve and monitor  the knot HST-1 at optical, radio and X-rays \citep{Biretta1999, Giroletti2012, Meyer2013, Snios2019, Thimmappa2024}. This knot is projected at 60 pc, 0\farcs8, from the nucleus, and the direct monitoring of it showed that it was brighter than the AGN core for many years. 

M87 is a low luminosity, radio loud FR I type AGN with a SMBH 30 times that of HE0435 \citep{EHT2019}. HE0435 is radio quiet, with a reported 5 GHz flux density of 113 $\mu$Jy \citet{Jackson2015}, (uncorrected for lensing magnification), 
which gives an f$_{5GHz}$/f$_{B}\approx$1/3 after extrapolating from the V magnitude data of \citet{Kochanek2006} assuming an optical slope of 1.  
A distinct component of extended X-ray emission may be the additional component required to explain the increase of the X-ray to optical luminosity ratio observed for radio-loud quasars \citep{Zamorani1981, Worrall1987, Wu2013}. Although HE0435 is radio-quiet, it has a 2 keV X-ray to 2500 
\AA\  logarithmic slope of $\alpha_{ox}=1.29$, that is similar to slopes for radio-loud AGN.

\citet{Jackson2015} compare the radio image of HE0435 to that predicted by various models of the source structure when folded through the gravitational lens model. They consider spherically symmetric models, and reject a point-like radio source, or an emitting region larger than 200 mas, with a preferred size of $\sim80$ pc.  An asymmetric X-ray source of that size and with flux of a few percent of the core quasar at the optical position could be measured to have a centroid position a few mas from the core. 

\vspace{0.15cm}
Moreover, models of GHz peaked and compact symmetric object (CSO) radio sources predict X-ray emission from sub-kpc-scale lobes and jets, even with a weak radio nucleus \citep{Stawarz2008, Migliori2014}. \citet{Krol2023} present evidence for such lobes of size 7 to 25 pc, based on broadband spectra of two nearby objects. 

\vspace{0.15cm}
In nearby type II AGN, extended hard continuum and Fe-K emission have been measured on scales of 100s of pc \citep{Marinucci2012, Marinucci2017, Fabbiano2017, Fabbiano2018a, Jones2020, Jones2021, TrindadeFalcao2023a, TrindadeFalcao2023b}. This may be due to outflowing energetic winds  \citep[e.g.,][]{Maksym2023}, or interaction of hard photons with extended galactic structures \citep[e.g.,][]{Feruglio2020, Fabbiano2018b}.

\vspace{0.15cm}

 Another possible (but less probable) explanation for the optical-X-ray offset, could be the presence of a pair of AGN. Such pairs are predicted to exist as an intermediate stage of galaxy mergers, but their occurrence at high redshift is still uncertain (see \citealt{DeRosa2019} and \citealt{Volonteri2021Nature} for extensive reviews).
 AGN pairs separated by tens of parsec are rare, only 0.1\% to 1\% of all quasars \citep{Volonteri2009,Kelley2019}. Moreover, to date there is a single precedent for a spatially resolved dual AGN at high redshift, where one member dominates in X-rays while the other dominates the optical light \citep{Chen2023}. Therefore, it would be remarkable if this lensed source were such a system.  Evidence for such a binary AGN could come from the observation of split emission lines (e.g., \citealt{Liu2018}), but these are not evident in spectra published for HE0435 \citep{Wisotzki2002, Melo2023}. At redshifts greater than $\approx1$ AGN pairs can only be directly imaged at projected separations larger than several kpc  (e.g., \citealt{Chen2022} and Fig. 7 of \citealt{Spingola2022} and references therein).
Therefore, HE0435 is unlikely such a binary AGN system. Nevertheless, the method presented in this work shows how the combination of high angular resolution optical/X-ray observations and gravitational lensing can potentially discover such pairs in large samples. 

 To what extent might errors in the high-fidelity, 2-dimensional point spread function that underlies our analysis cause the source displacement in Figure 5? We use the exact same simulation for all four image predictions and therefore expect the same putative bias in every image. Relative displacements due to such a bias would therefore cancel when we register the data to image A. Noting that lensing produces mirror images, images B and D  would have to be moved about 20 mas almost due east, while the same simulation would have to move image C 20 mas almost due south, all relative to a fixed position for image A. 

\subsection{X-ray variability over 14 years}\label{sec:xray_variability_discussion}

In Fig.~\ref{fig:fluxVtime} we show the \axaf\ X-ray flux of the lensed images of HE0435 as a function of time. The \axaf\ observations cover a period of $\sim14$ years (Table \ref{table:chandra_observations}), allowing us to potentially study the variability over months to years time scales.

There is a general (but not exact) agreement in the variability of the lensed images, as expected since they correspond to the same background quasar. The lack of an exact correspondence of the flux variations proves that the microlensing effect is present at different levels in all of the images. This is the reason why we could not use the flux ratios to further improve on the lens model.  We highlight, that the sub-arcsec angular resolution of \axaf\ is crucial to avoid erroneous conclusions about the overall source variability by spatially resolving the individual lensed images, revealing that they are affected by microlensing.  Nevertheless, it is worth noticing a broad coherence with the long-term optical monitoring observations from the COSMOGRAIL programme \citep{Bonvin2016}, including an increase in flux since 2014.
However, the lack of \axaf\ observations between 2014 and 2021 and the presence of microlensing do not allow us to perform a robust study on the correlation between the optical and X-ray emissions.

\section{Conclusions}

In this paper we presented a precise localization of the X-ray and optical emissions in the lensed AGN HE0435 at $z=1.693$. Assuming a simple (but precise) parametric lens model, we find a relative offset between the two emissions of $3.0\pm0.5$ mas ($26\pm4$ pc). This work demonstrates how powerful the lensing magnification is for resolving X-ray and optical emissions with unprecedented precision at high redshifts. 
Beyond a redshift of about $z=0.7$, strongly lensed sources are the only systems that can spatially resolve X-ray emissions on sub-kpc scales. The \axaf archive contains about 60 lensed quasars spanning redshifts up to $z\approx4$, which can be exploited to perform mas-scale X-ray astrometry, as shown here.  As a special case, we demonstrated that the varstrometry method can be used also at X-rays to distinguish pairs of emitting regions, like a SMBH and a knot in the jet or dual AGN systems.

In the future, this method can be applied to vastly more quadruply lensed AGNs, and at different wavelengths. The Vera C. Rubin Observatory and the Square Kilometer Array are due to come online in just a few years, and along with the \emph{Euclid} and \emph{Roman} space observatories are projected to find $\sim 10^5$ strong lensing systems \citep{Koopmans2004, McKean2015, Collett2015}. Among them, up to 10 \% will be AGN, and about 30\% will be lensed into four images. Therefore, there will be about $\sim3000$ quadruply imaged AGN available to apply the method presented here. 
\axaf\ has the capability to operate through those years. Alternately, only a future sub-arcsecond resolution X-ray telescope such as proposed for the \emph{Lynx} observatory \citep{Gaskin2018} can perform these measurements. A 2~ks \emph{Lynx} observation is comparable to a Chandra observation $\sim50$~ ks, allowing one to perform 1000 X-ray investigations of strongly lensed AGN in a 10-year program. 
With such a large number of lensing systems observed in the future, the numbers in a favorable configuration will likely be sufficient to investigate the relative location of X-ray, optical and radio emissions to constrain the occurrence and the origin of multi-wavelengths offsets across over cosmic time.

\begin{acknowledgments}
We thank A. Trindade Falcao and G. Fabbiano for discussion and references on outflows from type II AGN. We thank Giulia Migliori and J\'{u}lia Sisk-Reyn\'{e}s for comments on the manuscript. We thank the anonymous referee for useful suggestions.  D.A.S. and A.B. were supported by NASA contract NAS8-03060 to the \axaf\ X-ray Center, and by by CXC grant AR3-24007X. CS acknowledges financial support from the Italian National Institute for Astrophysics (INAF, FO: 1.05.12.04.04).
This paper is based in part on the Schiff Award undergraduate thesis of AR at Brandeis University. This work has made use of data from the European Space Agency (ESA) mission \emph{Gaia}(\url{https://www.cosmos.esa.int/gaia}), processed by the \emph{Gaia}
Data Processing and Analysis Consortium (DPAC, \url{https://www.cosmos.esa.int/web/gaia/dpac/consortium}). Funding for the DPAC
has been provided by national institutions, in particular the institutions participating in the \emph{Gaia} Multilateral Agreement. This research has made use of NASA's Astrophysics Data System Bibliographic Services, and of the NASA/IPAC Extragalactic Database, which is funded by the National Aeronautics and Space Administration and operated by the California Institute of Technology.
\end{acknowledgments}

\facility{Chandra(ACIS-S). We have used archival data from programs   	08700238 (PI: Schechter), 11700501 (PI: Kochanek), 14700473 (PI: Kochanek), and 22700407 (PI: Pooley). 
This research shows observations made with the NASA/ESA Hubble Space Telescope obtained from the Space Telescope Science Institute, which is operated by the Association of Universities for Research in Astronomy, Inc., under NASA contract NAS 5–26555. These observations are associated with program(s) 12889 (PI: Suyu), 13113 (PI: Kochanek) and 9744 (PI: Kochanek).
}

\software{gravlens \citep{Keeton2001}, ciao-4.13 \citep{Fruscione2006}, SAOTrace-2.0.4\_03 \citep{Jerius2004}, Marx-5.5.2 \citep{Davis2012}, SAOImageDS9 Version 8.4.1 \citep{Joye2003}, Astropy \citep{Astropy2013,Astropy2018,Astropy2022}.}

\vspace{5mm}

\appendix

\section{The astrometric analysis algorithm}

STEP 1: Use SAOTrace and Marx (with EDSER subpixel redistribution) to generate high fidelity simulations of point sources at the positions of the four X-ray images. This is done separately for each distinct observation. To run each simulation, we use the actual aspect solution and dither pattern of that observation.  We merge 1000 simulations, each using the actual X-ray flux and spectrum.  The simulations include pileup and the readout streak, although not noticable in the observations of HE0435. 

STEP 2: In an array perpendicular and parallel to the local caustic, take trial source plane positions for the X-ray emission as shown in Fig.~\ref{fig:CritPlot}. At each position, use the high-fidelity lens mass model from Section~\ref{sec:lensmodel} to predict the 4 image positions.

STEP 3: Construct a template for the predicted X-ray counts, placing the simulated point source images from Step 1 at the separations  predicted in Step 2.

STEP 4:  Bin the observed X-ray data into an array of pixels. We used 0\farcs246 pixels (1/2 ACIS pixel), in a 24 x 24 array. We observe n$_i$ counts in each pixel.

STEP 5: Raster the model from Step 3 in two dimensions, and re-sort into the bins of the observed data, to predict $\lambda_i$ counts per bin. Each simulated X-ray image has an arbitrary, uninteresting, renormalization parameter to account for flux ratio anomalies. Counts from all four images, plus the non-X-ray background, contribute to the 
$\lambda_i$  of each bin.

STEP 6: Compute the maximum likelihood
\begin{equation}
C=-2P =-2 \sum_{i=1}^{576}\ln(\lambda_i^{n_i} \exp(-\lambda_i)/n_i!).
\end{equation}

and vary the four image normalizations to minimize -1/2 C.

STEP 7: For each trial source position, interpolate within the raster to find the best registration of the predicted images. 

STEP 8: At each trial source position, compute the maximum likelihood,  (minimum C) of the best registration of the predicted images to the data.  Iterate for a range of trial source positions.

STEP 9: Define confidence regions for the X-ray source location, noting that the likelihood value differences from the minimum value are distributed as $\chi^2$ with 2 degrees of freedom.

\bibliography{he0435}{}
\bibliographystyle{aasjournal}

\end{document}